# The influence of electron beam irradiation on the chemical and structural properties of medical grade Polyurethane

**Sukyoung Shin\*, Soonhyouk Lee**

Ewha Medical Research Institute, School of Medicine, Ewha Womans University, Seoul 158-710, Republic of Korea

Thermo plastic polyurethane (TPU) provides excellent bio-compatibility, flexibility and good irradiation resistance; however, extremely high irradiation doses can alter the structure and function of macromolecules, resulting in oxidation, chain scission and cross-linking. In this study, the effects of e-beam irradiation on the medical grade thermo plastic polyurethane were studied. The changes in the chain length and their distribution as well as the changes in molecular structure were studied. The GPC (Gel Permeation Chromatography) results show that the oxidative decomposition is followed by a decrease in molecular mass together with an increase in polydispersity. This indicates a very inhomogeneous degradation, which is a consequence of the specific course and of the intensity of oxidative degradation. This was confirmed by means of mechanical property measurements. Overall, this study demonstrated that the medical grade TPU was affected by radiation exposure, particularly at high irradiation doses.

## I. INTRODUCTION

Thermo plastic polyurethane (TPU) elastomers are flexible biomaterials that have many end use applications, owing to its combination of excellent bio-stability, bio-compatibility, processability and desirable mechanical properties such as abrasion resistance, toughness, flexibility, durability and tensile strength [1]. TPUs are often referred to as segmented block copolymers as they consist of a hard and soft phase which either mix or segregate due to their immiscibility and produce phase mixed or phase separated morphologies. Each of the hard and soft segments is connected by means of urethane linkages, where the hard segment provides the physical crosslinks within the soft segment matrix [2,3]. Implant devices which contain such elastomers have significantly degraded in vivo after exposure to long-term biological environments as a result of hydrolytic or oxidative mechanisms [3-6]. Polyester TPUs are no longer used for devices that are required for long-term implantation due to poor hydrolytic stability. Polyether TPUs are hydrolytically stable yet they can undergo oxidative degradation in several forms including oxidation and environmental stress in the in vivo environment [1, 7, 8]. A common type of medical grade TPU is pellethane, which has been widely used as a biomaterial since its introduction in 1977 [9-11]. This material is employed to manufacture medical devices including those that are implanted and for this reason it is critical that the final product is sterilized before use. Conversely, such sterilization processes like dry steam, heat and high energy irradiation can have unfavorable effects on medical grade polymers such as extensive material degradation and plastic deformation [12-14]. In terms of high energy radio sterilization, the susceptibility of TPU to these processes with respect to crosslinking and degradation is highly dependent on the chemical. With regards to TPU material, different types of degradation processes can emerge subsequent to irradiation [15-18]. Furthermore, it is acknowledged that TPU elastomers can experience substantial structural changes when exposed to UV-irradiation which causes deterioration in their morphology[19]. There is little information in the literature about identifying the effects of irradiation conducted in an air atmosphere, on medical grade PU (e.g. pellethane 35D) or any systematic correlation between the segment composition and their resulting properties. The objective of this study is to quantify the effects of irradiation on the properties of TPU through GPC and mechanical property analysis.

## II. METHOD AND MATERIAL

Tubing was constructed with a 35D, 40D, 55D and 72D Thermo plastic polyurethane material. Elasthane and pellethane are polyether base thermoplastic polyurethanes from The Polymer Technology Group and Lubrizol respectively. Carbothane is a family of aliphatic and aromatic polycarbonate-based thermos plastic polyurethanes from Lubrizol. E-beam sterilization was performed by Nutek in Hayward, CA, using a 10 MeV accelerator. Samples were irradiated in a serial fashion via a conveyance system for approximately 10 minutes per dose.

Samples used for Molecular weight analysis were placed in the oven for accelerated aging simulation at 55C with 50% relative humidity. Table 1 summarized TPU material with different hardness grade used, e-beam condition and accelerated simulated aging time.

| TPU type | E-beam condition |
|---|---|
| Elasthane 35D | control |
| | e-beam @ 35kGy |
| | e-beam @ 50kGy |
| | e-beam @ 50kGy & aged 4wks |
| Carbothane 40D | control |
| | e-beam @ 35kGy |
| | e-beam @ 50kGy |
| | e-beam @ 50kGy & aged 4wks |
| Samples used for tensile test | |
| Pellethanee 55D | control |
| | e-beam @ 80kGy & aged 48 months |
| Pellethanee 55D | control |
| | e-beam @ 80kGy & aged 48 months |
| Pellethane 72D | control |
| | e-beam @ 80kGy & aged 48 months |

Table 1. Materials and e-beam radiation dose information.

The GPC system used for this work was calibrated using Agilent / Polymer Laboratories Easi Vial poly methyl meth acrylate (PMMA) calibrants. The highest molecular weight calibrant was considered to be 'excluded' and was not used in the calibration. The results are expressed as the 'PMMA equivalent' molecular weights and it should be appreciated that there could be considerable differences between these PMMA equivalents and the true molecular weights of the polymer. Data were analyzed to determine the average number molecular weight (Mn), weight average molecular weight (Mw) and polydispersity (Mw/Mn).

All materials were evaluated by mechanical testing. A Criterion Universal Testing Systems from MTS was used to mechanically test all materials in tensile mode. A 2/G MTS loadframe was used to mechanically test in tensile mode. For each sample condition, 10 specimen were tested and single average data point is reported in figures from 4 to 9.

## III. RESULTS AND DISCUSSION

### 1. GPC Molecular Weight analysis

Electron beam may be used on the material to induce effects such as chain scission (which makes the polymer chain shorter) and cross linking. The result is a change in the properties of the polymer which is intended to extend the range of applications for the material. The effect of the electron beam can cause the degradation of polymers, breaking chains and therefore reducing the molecular weight. Chain scission is the breaking apart of molecular chains to produce required molecular sub-units from the chain. Electron beam processing provides chain scission without the use of harsh chemicals usually utilized to initiate chain scission. Polymer can be cross-linked as well using high-energy ionizing radiation, i.e. electron beam (or e-beam, e beam), gamma, or x-ray. E-beam irradiation creates free radicals which will often chemically react in various ways, sometimes at slow reaction rates. The free radicals can recombine forming the crosslinks. The degree of crosslinking depends upon the polymer and radiation dose. One of the benefits of using irradiation for crosslinking is that the degree of crosslinking can be easily controlled by the amount of dose. Furthermore, oxidation can continue after irradiation causing changes in properties with time. Electron beam processing of thermoplastic material also results in an array of enhancements, such as an increase in tensile strength when polymers are cross-linked.

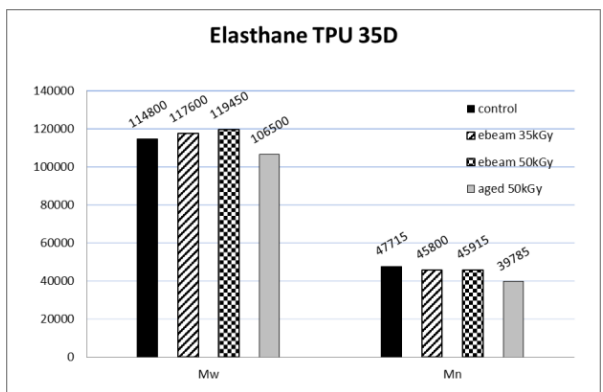

Figure 1. Molecular weight analysis summary of elasthane 35D.

Figure 1 shows GPC measurement of the molecular weight distribution of elasthane TPU 35D samples reveals increase in number average molecular weight. A conceivable reason for the molecular weight increased by the higher dosage was a crosslinking between free monomers or polymer chains. While the Mn, value of elasthane samples did not change. Increased MW, was attributed to crosslinking of the polymer, and may also reflect the loss of low molecular weight fragments. This suggests degradation, in the absence of stress, at the surface but not the bulk of the polymer.

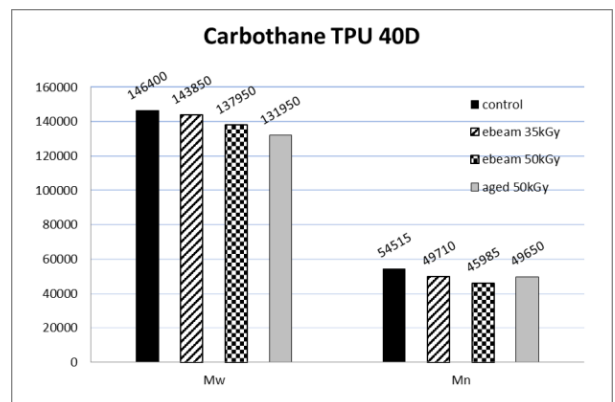

Figure 2. Molecular weight analysis summary of carbothane 40D.

For carbothane TPU 40D, there was a decrease in molecular weight numbers for higher irradiation doses, with exception for aged Mn, where the difference was not statistically significant. It can be explained that chain scission has prevailed for carbothane TPU material.

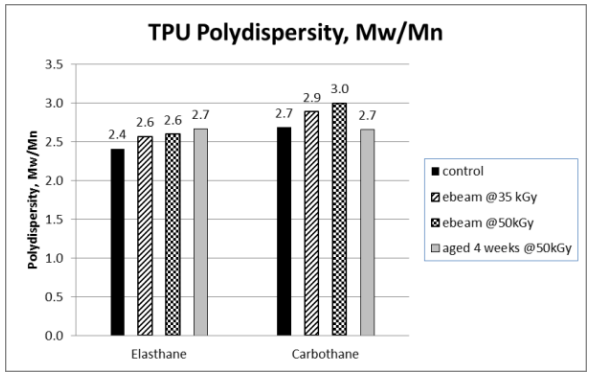

Figure 3. Polydispersity summary of Polyurethane post e-beam & ageing.

In Figure 3, the GPC results show that the oxidative decomposition is followed by a decrease in molecular mass together with an increase in polydispersity.

## 2. Mechanical Property analysis

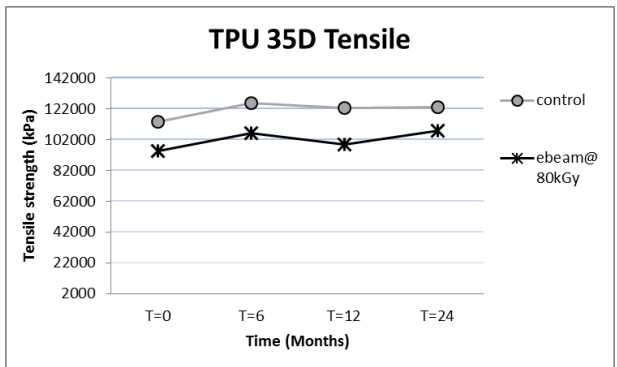

Figure 4. Tensile strength property summary of pellethane 35D post e-beam & ageing.

Figure 4 shows significant tensile differences were observed between pellethane 35D two-year E-beam results for time zero and time twelve months. Actual numerical differences between E-beam air groups were in line with differences observed between controls.

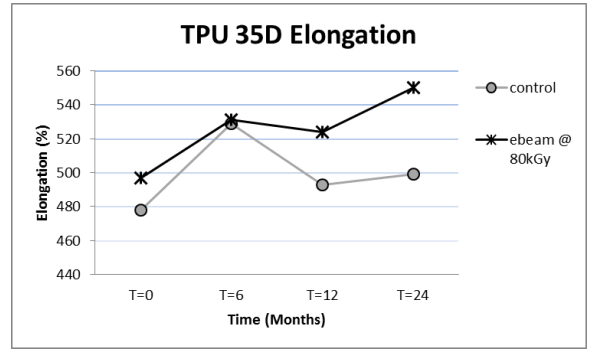

Figure 5. Elongation property summary of pellethane 35D post e-beam & ageing.

When compared to two-year control data, two-year E-beam showed a significant different elongation result. The numerical difference (499% versus 550%) is fairly modest and likely reflects no practical significance. A minimum elongation specification of 400% was available for pellethane 35D tubing. Materials from E-beam sterilization groups exceeded this specification.

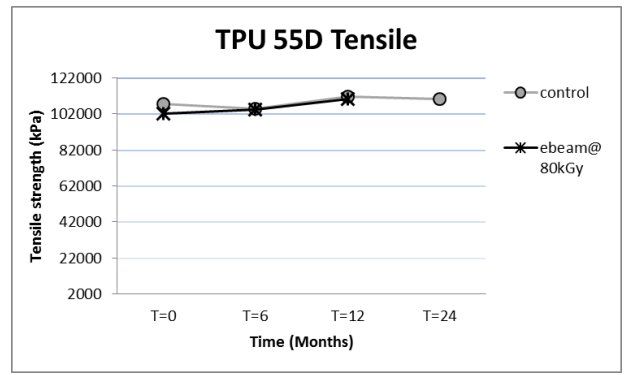

Figure 6. Tensile strength property summary of pellethane 55D post e-beam & ageing.

No significant differences were observed for pellethane 55D two-year E-beam data from other time points.

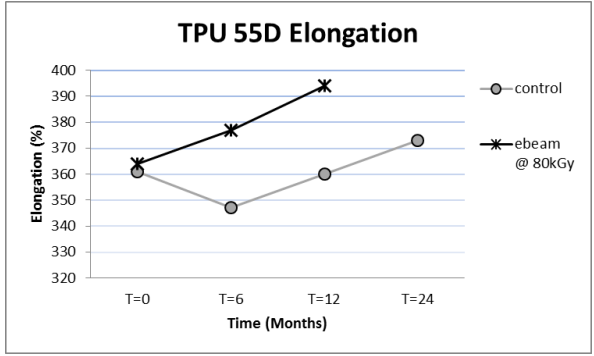

Figure 7. Elongation property summary of pellethane 55D post e-beam & ageing.

In figure 7, significant elongation differences were observed between pellethane 55D two-year E-beam materials from times zero and six. However, two-year E-beam data was limited to n=4 specimens and actual numerical differences were modest and in line with those observed between controls.

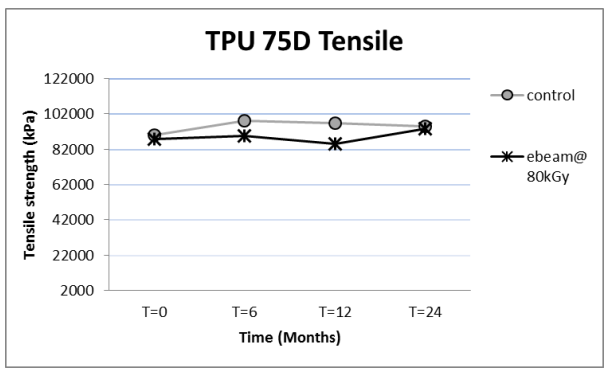

Figure 8. Tensile strength property summary of pellethane 75D post e-beam & ageing.

No significant differences were observed for pellethane 75D two-year E-beam data from other time points.

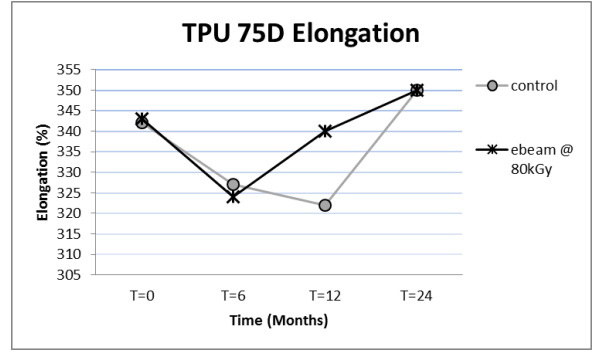

Figure 9. Elongation property summary of pellethane 75D post e-beam & ageing.

At two years, pellethane 75D E-beam elongation was statistically significantly different than respective two-year control elongation, but the actual numerical difference between two-year control and e-beam (350% versus 338%) was quite modest, reflecting no likely practical significance.

Overall, TPU degradation by high dose e-beam irradiation is not as severe as other medical grade elastomeric materials such as PEBA which reported fair amount of decrease in molecular weight and mechanical properties post irradiation.

IV. CONCLUSION

Irradiation of medical grade PU and aging effect while using a commercially available electron beam irradiator resulted in considerable modifications to the materials properties as analyzed GPC and tensile test. The material modifications triggered by irradiation exposure were analyzed by molecular weight change and mechanical properties. The GPC results show that the oxidative decomposition is followed by a decrease in molecular mass together with an increase in polydispersity. This indicates a very in-homogeneous degradation, which is a consequence of the specific course and of the intensity of oxidative degradation. This was confirmed by means of mechanical property measurements.

## ACKNOWLEDGEMENT

This work was supported by the Nuclear Power Core Technology Development Program of the Korea Institute of Energy Technology Evaluation and Planning (KETEP), granted financial resource from the Ministry of Trade, Industry & Energy, Republic of Korea. (No. 20131510400050)